\documentstyle[12pt,a4wide,cite,rotate,epsf]{article}

\newcommand{\wh}{\widehat}
\newcommand{\gf}{\gamma_5}
\newcommand{\IM}{\mbox{\rm Im}}
\newcommand{\nn}{\nonumber}

\newcommand{\wt}{\widetilde}
\newcommand{\al}{\alpha_{s}}
\newcommand{\ds}{\Delta S=2}
\newcommand{\ve}{\varepsilon}
\newcommand{\eqn}[1]{(\ref{#1})}

\newcommand{\ap}{\frac{\al}{\pi}}
\newcommand{\als}{\alpha_{s}(s)}
\newcommand{\aps}{\frac{\als}{\pi}}
\newcommand{\MSb}{{\overline{MS}}}
\newcommand{\tvs}{\vbox{\vskip 6mm}}
\newcommand{\smvs}{\vbox{\vskip 8mm}}

\newcommand{\La}{\Lambda_{\overline{MS}}}

\newcommand{\lsim}{~{}_{\textstyle\sim}^{\textstyle <}~}
\newcommand{\newsection}[1]{\section{#1}\setcounter{equation}{0}}

\newcommand{\be}{\begin{equation}}
\newcommand{\ee}{\end{equation}}
\newcommand{\ba}{\begin{array}{c}}
\newcommand{\ea}{\end{array}}
\newcommand{\beqn}{\begin{eqnarray}}
\newcommand{\eeqn}{\end{eqnarray}}

\newcommand{\bi}{\begin{itemize}}
\newcommand{\ei}{\end{itemize}}

\newcommand{\cO}{{\cal O}}

\newcommand{\rms}{\rm\scriptsize}

\begin{document}
\bibliographystyle{physics}


\begin{titlepage}
\begin{flushright}
{CERN--TH.7151/94}\\
{FTUV/94--6}\\
{IFIC/94--3}
\end{flushright}
\vspace*{5mm}
\begin{center}
{\LARGE\sf QCD corrections to inclusive $\Delta S=1,2$ transitions\\
\vspace*{4mm}
at the next-to-leading order}
\vspace*{10mm}\\
{\normalsize\bf Matthias Jamin}\\
{\small\sl Theory Division, CERN, CH--1211 Geneva 23, Switzerland}\\
and \\
{\normalsize\bf Antonio Pich}\\
{\small\sl Departament de F\'\i sica Te\`orica, Universitat de Val\`encia}\\
{\small\sl and IFIC, Centre Mixte Universitat de Val\`encia -- CSIC}\\
{\small\sl E--46100 Burjassot, Val\`encia, Spain}
\vspace{12mm}\\
{\bf Abstract}
\end{center}
\noindent
The 2--point functions for $\Delta S=1$ current-current and QCD-penguin
operators, as well as for the $\ds$ operator, are calculated at the
next-to-leading order. The calculation is performed in two different
renormalization schemes for $\gamma_5$, and the compatibility of the
results obtained in the two schemes is verified. The scale- and
scheme-invariant combinations of spectral 2--point functions and
corresponding Wilson-coefficients are constructed and analyzed. For
$\Delta S=1$, the QCD corrections to the CP-conserving part, dominated
by current-current operators, are 40\%--120\% at
$q^2=\left(1-3\,\mbox{\rm GeV}\right)^2$, whereas the correction to the
imaginary part, mainly coming from the penguin operator $Q_6$, are
100\%--240\%. The large size of the gluonic corrections to current-current
operators provides a qualitative understanding of the observed enhancement
of $\Delta I=1/2$ transitions. In the $\Delta S=2$ sector the QCD
corrections are quite moderate ($\approx -20\%$).
\vfill
\begin{flushleft}
{CERN--TH.7151/94}\\
{February 1994}
\end{flushleft}

\end{titlepage}

\newpage
\setcounter{page}{1}


\newsection{Introduction}

Recent years have witnessed considerable improvement in Standard Model
calculations of non-leptonic weak decays. In particular, the effective
Hamiltonians for flavour-changing $\Delta F=1$
\cite{ACMP:81,BJLW:92a,CFMR:93a,BJL:93} as well as
$\Delta F=2$ \cite{BJW:90,HN:93} transitions were calculated at the
next-to-leading order (NLO) in renormalization group (RG) improved
perturbation theory. The necessary computations included the determination
of 2--loop anomalous dimension matrices for current-current, QCD-penguin,
and electroweak penguin operators
\cite{ACMP:81,BJLW:92a,BW:90,BJLW:92b,BJL:93a,CR:93,CFMR:93b}.
The benefit from such calculations is the following:
\begin{itemize}
\item the size of the NLO short distance corrections to the coefficient
 functions was obtained,
\item this allowed an estimation of the scale down to which RG evolution
 is possible, before perturbation theory breaks down, to be around
 $1\,\mbox{\rm GeV}$,
\item only a NLO calculation allows for a meaningful use of the QCD scale
 $\La$, extracted for example in deep inelastic scattering,
 $\tau$ decay or LEP
 experiments,
\item and it made possible to attack the question of the dependence of
 coefficient functions on the renormalization scheme, e.g. the definition
 of $\gamma_5$ in an arbitrary space-time dimension, which first appears
 at the next-to-leading order.
\end{itemize}

Sadly enough, this is not the whole story. In order to fully calculate
the decay amplitude for a certain process, we also need to know the
matrix elements of the operators appearing in the effective Hamiltonian,
between the hadronic initial and final states. This part is much more
difficult, for it involves non-perturbative dynamics at low energies.
Methods to attempt this involved task include lattice gauge theory
\cite{GKS:92,BER:93,SAC:93}, $1/N$-expansion \cite{BBG:87,BBG:87a,BBG:88},
chiral perturbation theory \cite{KMW:91,KDHMW:92}, QCD sum rules
\cite{PD:85,GPD:85,PGD:86,PD:87,PI:89,PDPPR:91,PD:91}, and mixed approaches
involving functional integration of quark fields \cite{PD:91}. A strategy
to obtain the matrix elements for $\Delta S=1$ decays at NLO as far as
possible from experimental data was also advocated in ref.~\cite{BJL:93}.

All these methods suffer from more or less severe drawbacks.
Although the lattice could eventually become the ultimate tool for the
calculation of matrix elements, the  precision of present lattice
results is still very poor. $1/N$-expansion and chiral perturbation
theory are not yet directly related to the fundamental QCD Lagrangian,
and therefore, a sound matching of matrix elements calculated in one of
these methods and the coefficient functions, obtained in perturbation
theory, is still not possible. The bridge between the effective low-energy
chiral Lagrangian and the underlying QCD theory can be built with
functional bosonization techniques \cite{PD:91}, or using QCD sum rules
to relate the two energy regimes
\cite{PD:85,GPD:85,PGD:86,PD:87,PI:89,PDPPR:91,PD:91}. Unfortunately, the
present determinations are not very accurate. Finally, there is not
enough experimental information to obtain all matrix elements, say
for $K\rightarrow\pi\pi$ decays, from the phenomenological method of
ref.~\cite{BJL:93}.

The problem becomes much easier at the inclusive level, where the
properties of the non-leptonic effective weak Hamiltonian can be analyzed
within QCD \cite{PD:85,GPD:85,PGD:86,PD:87,PI:89,PD:91}.
Given, for instance, the short-distance $\Delta S=1$ Hamiltonian
\cite{BJLW:92a,BJL:93}
\begin{equation}
\label{eq:hamiltonian}
{\cal H}^{\Delta S=1}_{\mbox{\rms eff}} \; = \;
{G_F\over\sqrt{2}}\,V_{ud}^{\phantom{*}} V_{us}^{*} \sum_{i} C_{i}(\mu^2)
   \, Q_{i} \,,
\end{equation}
obtained through the operator product
expansion, one considers the 2--point function
\begin{eqnarray}
\label{eq:correlator}
\Psi^{\Delta S=1}(q^2) & \equiv & i \int \! dx \, e^{iqx} \,
\big<0\vert \, T\{\,{\cal H}^{\Delta S=1}_{\mbox{\rms eff}}(x)\,
{\cal H}^{\Delta S=1}_{\mbox{\rms eff}}(0)^\dagger\}\vert0\big>
\nonumber \\
& = & \left({G_F\over\sqrt{2}}\right)^2
\left| V_{ud}^{\phantom{*}} V_{us}^{*}\right|^2\,
\sum_{i,j} \, C_i(\mu^2) \,
C_j^*(\mu^2) \,\Psi_{ij}(q^2) \,.
\end{eqnarray}
This vacuum-to-vacuum
correlator can be studied with perturbative QCD methods, allowing for a
consistent combination of Wilson-coefficients $C_i(\mu^2)$ and 2--point
functions of the 4--quark operators, $\Psi_{ij}$, in such a way that
the renormalization scheme and scale dependences exactly cancel (to
the computed order). The associated spectral function
$\frac{1}{\pi}\mbox{\rm Im}\Psi^{\Delta S=1}(q^2)$
is a quantity with definite physical information.
It describes in an inclusive way how the weak Hamiltonian couples
the vacuum to physical states of a given invariant mass.
General properties like the observed enhancement of $\Delta I=1/2$
transitions can be then rigorously analyzed at the inclusive level.

A detailed analysis of 2--point functions associated with $\Delta S=1$
and $\Delta S=2$ operators was presented in ref. \cite{PD:91}, where
the $\cO(\alpha_s)$ corrections to the corresponding correlators
$\Psi_{ij}$ were calculated. The NLO corrections to the $\Delta I=1/2\, $
2--point functions were found to be very large \cite{PD:91}, confirming
the QCD enhancement obtained in a previous approximate calculation
\cite{PI:89}. The results of ref.~\cite{PD:91} were, however, incomplete
because the NLO corrections to the Wilson-coefficients of penguin
operators were still missing.
With the progress achieved for the Wilson-coefficient functions mentioned
above, we are now in a position to match matrix elements and coefficient
functions consistently at NLO.

To get a sensible result, we obviously need to use the same
renormalization scheme conventions on both sides of the calculation.
Unfortunately, for technical reasons, different bases of operators have
been used in the 2--point function and Wilson-coefficient calculations.
In order to avoid ambiguities coming from the definition of $\gamma_5$
in $d\not=4$ dimensions, a set of colour-singlet 4--quark operators was
used in ref. \cite{PD:91} to perform the 2--point-function calculation;
the computation was done with dimensional regularization and a naively
anticommuting $\gamma_5$ (NDR scheme). While that is fine in the
leading logarithmic approximation, the basis of colour-singlet operators
does not close under renormalization at the NLO. The Fierz-transformations,
which are needed to relate some of the operators in the process of
renormalization, are broken by $\cO(\alpha_s)$ corrections, and additional
contributions have to be taken into account. For this reason, we shall
reconsider the calculation of the 2-point functions of 4--quark operators
in this work.

We shall use the same basis of operators which has been taken for
the calculation of the Wilson-coefficients, so that we can directly
incorporate the results of refs. \cite{BJLW:92a,BW:90,BJLW:92b,CFMR:93b}.
The presence of colour-non-singlet operators in this basis gives rise to
$\gamma_5$ complications in the 2--point-function evaluation.
Like for the calculation of the anomalous dimension matrices in refs.
\cite{BJLW:92a,BW:90,BJLW:92b,CFMR:93b}, we shall perform the calculation
in two different schemes for $\gamma_5$, to have explicit tests on our
result. A direct computation with a naively anticommuting $\gamma_5$
is {\em not} possible, since two diagrams include traces
of odd numbers of $\gamma_5$; but we shall show, how nevertheless a result
in the NDR scheme can be obtained. For the second computation, the consistent
definition of a non-anticommuting $\gamma_5$ in arbitrary dimensions
according to 't Hooft and Veltman (HV~scheme) \cite{HV:72,BM:77,BW:90}
is used.

In sect.~2, we shall discuss the general structure of 2--point functions
of 4--quark operators. As a first step towards the explicit calculation
for the $\Delta S=1$ case, in sect.~3, the 2--point functions of
current-current operators are computed, and the full set including
QCD-penguins is presented in sect.~4. A numerical analysis of the results
is given in sect.~5. In sect.~6, we evaluate the 2--point function
for the $\ds$ operators. A comparison with the results of
ref.~\cite{PD:91}, together with some concluding remarks, is finally
given in sect.~7.

\newsection{General structure}

As a first step, let us discuss the general structure of the 2--point
functions of 4--quark operators and their renormalization. The bare
2--point function $\Psi^B(q^2)$ is defined by
\begin{equation}
\Psi^B(q^2) \; \equiv \; i \int \! dx \, e^{iqx} \, \big<0\vert \,
T\{\,Q^B(x)\,Q^{B^\dagger}(0)\}\vert0\big> \,.
\label{eq:2.1}
\end{equation}
$Q(x)$ can either be a single operator, or a vector of 4--quark operators,
in which case $\Psi^B(q^2)$ is a symmetric matrix\footnote{For this general
discussion, we shall assume $\Psi^B(q^2)$ to be a matrix.}.

Keeping only relevant terms up to next-to-leading order in $\al$,
and working with dimensional regularization, the regularized, but
yet unrenormalized 2--point function $\Psi^R(q^2,\mu^2)$ has an expansion
in $\ve$ ($d=4+2\ve$),
\begin{eqnarray}
\Psi^R(q^2,\mu^2)&=&-\,\frac{(q^2)^4}{(4\pi)^6}\,\biggl\{\,\biggl(\frac{-q^2}
{\mu^2}\biggr)^{3\ve}\frac{1}{3\ve}\,\Big[\,A+B\ve+\ldots\Big] \nn \\
\smvs
& & \hspace{7.7mm} +\,\ap\biggl(\frac{-q^2}{\mu^2}\biggr)^{4\ve}
\frac{1}{4\ve^2}\,\Big[\,C+D\ve+\ldots\Big] + {\cal O}(\al^2)\,\biggr\} \,.
\label{eq:2.2}
\end{eqnarray}
$\mu^2$ is a renormalization scale in the $\MSb$ scheme, that is, we have
redefined the scale of dimensional regularization $\nu^2$ to be
$\exp(\gamma_E)\,\mu^2/(4\pi)$, where $\gamma_E$ is Euler's constant, so
that only poles in $\ve$ have to be subtracted.
Our main goal will be to calculate the four matrices $A$, $B$, $C$, and $D$.

The renormalized 2--point function is given by
\begin{equation}
\Psi(q^2,\mu^2) \;=\; R_\MSb\,\Big[\,Z^{-1}\,\Psi^R(q^2,\mu^2)\,\big(Z^{-1}
\big)^T\,\Big] \,,
\label{eq:2.3}
\end{equation}
where $Z$ is the renormalization matrix of the 4--quark operators,
$Q^B\equiv Z\,Q$, and $R_\MSb$ means that additional poles in $\ve$,
stemming from the operator product, need to be subtracted.
In the minimal subtraction scheme, $Z$ has the general expansion
\begin{equation}
Z \; = \; 1 + \sum_{k=1}^{\infty} \Big(\ap\Big)^{k}\sum_{n=1}^{k}
\frac{Z_n^{(k)}}{\ve^n} \,.
\label{eq:2.4}
\end{equation}
The anomalous dimension matrix of 4--quark operators is defined through
\begin{equation}
\gamma \; \equiv \; Z^{-1}\mu\,\frac{dZ}{d\mu}\,\biggr\vert_{\ve=0}
\; = \; \gamma^{(1)}\,\ap + \gamma^{(2)}\,\Big(\ap\Big)^2 + \ldots \,.
\label{eq:2.5}
\end{equation}
Inserting the expansion for $Z$, eq.~\eqn{eq:2.4}, we find
\begin{equation}
\gamma^{(1)} \; = \; 2\,Z_1^{(1)} \,, \qquad
\gamma^{(2)} \; = \; 4\,Z_1^{(2)} \,, \qquad
Z_2^{(2)} \; = \; \frac{1}{4}\,Z_1^{(1)}\Big(\,2\,Z_1^{(1)}-\beta_1\,\Big)\,.
\label{eq:2.6}
\end{equation}
Here, $\beta_1=-\,(11N-2f)/6$ is the leading coefficient of the
$\beta$--function, $N$ and $f$ being the number of colours and flavours
respectively. For the operators which mediate $\Delta S=1$ transitions,
and shall be our interest for most part of the paper, the leading order
anomalous dimension matrix is known already since a long time
\cite{GL:74,AM:74,VZS:77,GW:79,GP:80}. However, the
full next-to-leading order matrix has only been obtained recently
\cite{ACMP:81,BJLW:92a,BW:90,BJLW:92b,CFMR:93b}.
To make contact with the notation of refs.~\cite{BJLW:92a,BJLW:92b},
we note that
\begin{equation}
\gamma^{(1)} \; = \; \frac{1}{4}\,\gamma_{BJLW}^{(0)} \,,
\qquad \hbox{and} \qquad
\gamma^{(2)} \; = \; \frac{1}{16}\,\gamma_{BJLW}^{(1)} \,.
\label{eq:2.6a}
\end{equation}

Using the results of eq.~\eqn{eq:2.6}, together with eq.~\eqn{eq:2.4}, up
to non-logarithmic corrections the renormalized 2--point function turns
out to be
\begin{equation}
\Psi(q^2,\mu^2) \; = \; -\,\frac{(q^2)^4}{(4\pi)^6}\,\biggl\{\,A\,L+
\ap\,\biggl[\,\frac{1}{2}\,CL^2+XL\,\biggr]\,\biggr\} \,,
\label{eq:2.7}
\end{equation}
where $L=\ln(-q^2/\mu^2)$, and
\begin{equation}
X \; = \; D-\frac{1}{2}\,\Big(\,\gamma^{(1)}B+B\,\gamma^{(1)^T}\,\Big) \,.
\label{eq:2.8}
\end{equation}
Because of renormalizability, it also follows that
\begin{equation}
C \; = \; \frac{1}{2}\,\Big(\,\gamma^{(1)}A+A\,\gamma^{(1)^T}\,\Big) \,.
\label{eq:2.9}
\end{equation}

For the rest of this work, we shall only be concerned with the spectral
function (the imaginary part of the 2--point function) which is
directly related to physical quantities:
\begin{equation}
\Phi(s,\mu^2) \; \equiv \; \frac{1}{\pi}\,\IM\,\Psi(q^2,\mu^2) \; = \;
\theta(s)\,
\frac{s^4}{(4\pi)^6}\,\biggl\{\,A+\ap\,\biggl[\,
C  \ln{\biggl|\frac{s}{\mu^2}\biggr|}
+X\,\biggr]\,\biggr\} \,,
\label{eq:2.10}
\end{equation}
with $s\equiv q^2$.
It is straightforward to see that $\Phi(s,\mu^2)$ satisfies a
homogeneous renormalization group equation (RGE),
\begin{equation}
\mu\,\frac{d}{d\mu}\,\Phi(s,\mu^2)+\gamma\,\Phi+\Phi\,\gamma^T \; = \; 0 \,.
\label{eq:2.11}
\end{equation}
This implies that
\begin{equation}
\wh\Phi(s) \; \equiv \; C^T(\mu^2)\,\Phi(s,\mu^2)\,C^*(\mu^2)
\label{eq:2.12}
\end{equation}
is a renormalization group invariant quantity, with $C(\mu^2)$ being the
Wilson-coefficient function of the 4--quark operators, which satisfies
the RGE
\begin{equation}
\biggl\{\,\mu\,\frac{d}{d\mu}-\gamma^T\,\biggr\}\,C(\mu^2) \; = \; 0 \,.
\label{eq:2.13}
\end{equation}
At the next-to-leading order, the coefficient function for $\Delta S=1$
operators can be found in refs.~\cite{BJLW:92a,CFMR:93a,BJL:93}.
The scale- and scheme-independence of $\wh\Phi$ should be clear, because
this function is just proportional to the physical spectral function
${1\over\pi}\mbox{\rm Im}\Psi^{\Delta S=1}(s)$. The scheme independence of
${\cal H}^{\Delta S=1}_{\mbox{\rms eff}}$ is carried over to the
2--point function.

We can easily sum up the next-to-leading logarithms in the 2--point function
by setting $\mu^2=s$, yielding
\begin{equation}
\Phi(s) \; = \; \theta(s)\,
\frac{s^4}{(4\pi)^6}\,\biggl\{\,A+\aps\,X\;\biggr\}\,,
\quad \hbox{and} \quad
\wh\Phi(s) \; = \; C^T(s)\,\Phi(s)\,C^*(s) \,.
\label{eq:2.14}
\end{equation}

Like the Wilson-coefficient function, also the spectral function at the
next-to-leading order, in particular the matrices $B$ and $D$,
depend on the renormalization scheme. If the renormalization matrices in
two schemes $Z_a$ and $Z_b$ are related by a finite shift,
\begin{equation}
Z_a \; = \; Z_b\,\biggl[\,1+\ap\,\Delta r\,\biggr] \,,
\label{eq:2.15}
\end{equation}
we find the following relation between $X$ in the two schemes,
\begin{equation}
X_b \; = \; X_a + \Delta r\,A+A\,\Delta r^T \,.
\label{eq:2.16}
\end{equation}
Together with the scheme dependence of the Wilson-coefficient functions
(eq.~(3.6) of ref.~\cite{BJL:93}),
\begin{equation}
C_b(\mu^2) \; = \; \biggl[\,1-\ap\,\Delta r^T\,\biggr]\,C_a(\mu^2) \,,
\label{eq:2.17}
\end{equation}
it is a trivial check that $\wh\Phi(s)$ is indeed scheme-independent
up to ${\cal O}(\al^2)$.

\newsection{Current-current operators}

As an introductory example, we shall first calculate the 2--point function
of the $\Delta S=1$ current-current operators, before embarking on the
full set including penguins:
\begin{equation}
Q_{1} \; = \; \left( \bar s_{\alpha} u_{\beta}  \right)_{\rm V-A}
              \left( \bar u_{\beta}  d_{\alpha} \right)_{\rm V-A} \,,
\qquad
Q_{2} \; = \; \left( \bar s u \right)_{\rm V-A}
              \left( \bar u d \right)_{\rm V-A} \,,
\label{eq:3.1}
\end{equation}
where $\alpha$, $\beta$ denote colour indices ($\alpha,\beta
=1,\ldots,N$) and the colour indices have been omitted for the
colour singlet operator $Q_2$. $(V-A)$ refers to $\gamma_{\mu} (1-\gf)$.
This basis closes under renormalization if penguin operators are
neglected.\footnote{For the HV scheme, $\gamma_\mu$ has to be taken
in 4 dimensions.}

In the course of the calculation, it will become useful to also study
2--point functions of the Fierz-transformed operators
\begin{equation}
\widetilde{Q}_1 \; = \; \left(\bar s d \right)_{\rm V-A}
                        \left(\bar u u \right)_{\rm V-A} \,,
\qquad
\widetilde{Q}_2 \; = \; \left(\bar s_\alpha d_\beta  \right)_{\rm V-A}
                        \left(\bar u_\beta  u_\alpha \right)_{\rm V-A} \,,
\label{eq:3.2}
\end{equation}
and mixtures of the bases \eqn{eq:3.1} and \eqn{eq:3.2}. Since these
mixtures do not close under renormalization, we will have to include
evanescent operators. This will be discussed in detail below.

The calculation of the 2--point function requires the evaluation of
the leading order 3--loop diagrams of fig.~1 and the next-to-leading
4--loop diagrams of fig.~2. The results of this evaluation are
summarized in tables~\ref{tab:1} and \ref{tab:2}, and will be
discussed in great detail in the following. A straightforward inspection
reveals that the topology 2e contains traces with an odd number of
$\gamma_5$'s. Actually, there are two diagrams of type 2e:
one with the fermion lines in the upper and lower loop circulating in
opposite directions, denoted by 2e, and one with the same direction,
denoted by 2e'. These {\em cannot} directly be calculated in renormalization
schemes with a naively anticommuting $\gamma_5$. For this reason,
and also to make direct contact with the NLO calculation of the
Wilson-coefficient function~\cite{BJLW:92a,CFMR:93a,BJL:93}, we shall
perform the calculation with a non-anticommuting $\gamma_5$, originally
due to 't Hooft and Veltman \cite{HV:72,BM:77,BW:90}, and
in addition we present a way to nevertheless obtain results in the
NDR scheme.

%
%
\begin{table}[thb]
\caption{Results for the lowest-order diagrams of fig.~1. \label{tab:1}}
\begin{center}
\renewcommand{\arraystretch}{1.4}
\begin{tabular}{|c|rr|}
\hline
Diagram & 1a & 1b \\
\hline
$A$ & $\frac{4}{45}$ & $\frac{4}{45}$
\\
$B^{NDR}$ & $-\frac{653}{675}$ & $-\frac{593}{675}$
\\
$B^{HV}$ & $-\frac{1637}{1575}$ & $-\frac{1637}{1575}$
\\
\hline
\end{tabular}
\renewcommand{\arraystretch}{1}
\end{center}
\end{table}
%

%
%
\begin{table}[thb]
\caption{Results for the $\cO(\al)$ diagrams of fig.~2
(Feynman gauge). \label{tab:2}}
\begin{center}
\renewcommand{\arraystretch}{1.4}
\begin{tabular}{|c|rrrrrrrrrr|}
\hline
Diagram & 2a & 2b & 2c & 2d & 2e & 2e' & 2f & 2g & 2g' & 2g''
\\
\hline
$C$ &
$\frac{1}{45}$ & $\frac{1}{45}$ & $-\frac{2}{45}$ &
$-\frac{2}{45}$ & $\frac{8}{45}$ & $-\frac{2}{45}$ &
$\frac{8}{45}$ & $\frac{4}{135}$ & $\frac{4}{135}$ & $\frac{4}{135}$ \\
$D^{NDR}$ &
$-\frac{61}{180}$ & $-\frac{19}{60}$ & $\frac{67}{90}$ & $\frac{7}{10}$ &
$-\frac{26}{9}$ & $\frac{67}{90}$ & $-\frac{122}{45}$ & $-\frac{31}{81}$ &
$-\frac{167}{405}$ & $-\frac{179}{405}$ \\
$D^{HV}$ &
$-\frac{1349}{3780}$ & $-\frac{1349}{3780}$ & $\frac{1643}{1890}$ &
$\frac{1643}{1890}$ & $-\frac{2866}{945}$ & $\frac{1643}{1890}$ &
$-\frac{2866}{945}$ & $-\frac{263}{567}$ & $-\frac{263}{567}$ &
$-\frac{263}{567}$ \\
\hline
\end{tabular}
\renewcommand{\arraystretch}{1}
\end{center}
\end{table}

\subsection{Current-current operators in the HV scheme}

Since the calculation is more transparent in the HV scheme, let us
begin with this case. We shall denote with $\Psi_{ij}$ a matrix
element of the general matrix $\Psi$, eq.~\eqn{eq:2.2}, corresponding
to the 2--point function of the operators $Q_i$ and $Q_j$. Then the
three entries for the $2\times2$ matrix for $Q_1$ and $Q_2$ (recall
that $\Psi$ is symmetric) are given by
\begin{eqnarray}
\Psi_{11} & = & N^2 \Psi_{1a}+N^2 C_f\Big[\,4\Psi_{2a}+2\Psi_{2e'}\,
\Big] \,, \label{eq:3.3} \\
\Psi_{12} & = & N \Psi_{1a}+N C_f\Big[\,4\Psi_{2a}+2\Psi_{2c}+
2\Psi_{2e}+2\Psi_{2e'}\,\Big] \,, \label{eq:3.4} \\
\Psi_{22} & = & N^2 \Psi_{1a}+N^2 C_f\Big[\,4\Psi_{2a}+2\Psi_{2c}\,
\Big]
 + N C_f \Psi_{2g} \,,
\label{eq:3.5}
\end{eqnarray}
where $C_f=(N^2-1)/2N$.
The contributions to the 2--point function from a given diagram,
$\Psi_{\mbox{\rms diag}}$, can be obtained by inserting into
eq.~\eqn{eq:2.2} the relevant entries of tables~\ref{tab:1} and \ref{tab:2}.
Including the corresponding colour factors and multiplicities which can be
read off from eqs.~\eqn{eq:3.3}--\eqn{eq:3.5}, we obtain the $2\times2$
matrices $A$, $B^{HV}$, $C$, and $D^{HV}$:
\begin{equation}
A \; = \; \frac{4}{45}\,N
\left( \begin{array}{cc}
N & 1 \\ 1 & N
\end{array} \right) \,, \qquad
B^{HV} \; = \; -\,\frac{1637}{140}\,A \,,
\label{eq:3.8}
\end{equation}
\begin{equation}
C \; = \; \frac{4}{15}\,N C_f
\left( \begin{array}{cc}
0 & 1 \\ 1 & 0
\end{array} \right) \,, \qquad
D^{HV} \; = \; \frac{1}{315}\,N C_f
\left( \begin{array}{cc}
42N & -1321 \\ -1321 & 42N
\end{array} \right) \,.
\label{eq:3.9}
\end{equation}
Inserting these results into eq.~\eqn{eq:2.8}, we arrive at
\begin{equation}
X^{HV} \; = \; \frac{2}{225}\,N C_f
\left( \begin{array}{cc}
15N & -121 \\ -121 & 15N
\end{array} \right) \,.
\label{eq:3.10}
\end{equation}

Although at this stage a statement about the size of the radiative
corrections is scheme-dependent, let us nevertheless perform this
exercise. Taking $\alpha_s(s)/\pi\approx 0.1$, from eq.~\eqn{eq:2.14}
we find a moderate 20\% correction in the diagonal, but the
off-diagonal terms are almost a factor of 2 compared with the
leading term. This already gives an indication of huge radiative
corrections in the final result. However, note that these contributions
are subleading in an expansion in $1/N$.

Several technical remarks on the calculation so far are in order:

i) Up to now, we only dealt with the current-current operators $Q_1$
and $Q_2$. In this case, the penguin type contribution of diagram 2g
in eq.~\eqn{eq:3.5} has been omitted for consistency. It will be taken
into account in the full result including penguin operators.

ii) Naively, the HV scheme breaks some Ward-identities, e.g., the weak
current is not conserved. We can enforce conservation of the weak
current by performing a finite renormalization which results in a
shift for $D^{HV}$. This shift is given by $-2C_f A$, and has been
incorporated\footnote{See also the discussion in
refs.~\cite{BJLW:92a,BJL:93,BJLW:92b}.}
into eq.~\eqn{eq:3.9}.

iii) In the course of the calculation of the anomalous dimension matrix
for 4--quark operators \cite{BW:90,BJLW:92b,CFMR:93b}, tensor
structures with six $\gamma$-matrices appear which have to be
projected onto the physical subset of operators. One example for the
projections used in refs.~\cite{BW:90,BJLW:92b} is
\begin{equation}
\gamma_\mu\gamma_\nu\gamma_\lambda(1-\gamma_5)\otimes
\gamma^\mu\gamma^\nu\gamma^\lambda(1-\gamma_5) \quad\longrightarrow\quad
4\,(4-\ve)\,\gamma_\mu(1-\gamma_5)\otimes\gamma^\mu(1-\gamma_5) \,.
\label{eq:3.11}
\end{equation}
Generally speaking, at $\cal O(\ve)$ this projection is arbitrary and a
specification of the projection has to be added to the definition of the
renormalization scheme. The projections used in refs.~\cite{BW:90,BJLW:92b}
have been chosen such that Fierz-relations in the current-current sector
are preserved. This were not the case for an arbitrary projection.
As a check, we have also performed the calculation with an arbitrary
projection, and have verified that scheme-invariant quantities are
indeed independent of this choice, as they should.

Now, the 2--point functions have to be calculated in accord with the
calculation of the anomalous dimensions. This means, we first have to
calculate the radiative correction to either of the operators, then perform
the projection onto the physical basis, and finally insert the resulting
expression into the 2--point function. The only place where this
treatment gives a different result, compared to a naive evaluation
of the 2--point function (given the above choice of projection),
is in diagrams 2e and 2f in the HV scheme. In the NDR scheme the
naive calculation immediately yields the correct result for all diagrams
except for the problems with $\gamma_5$ in 2e and 2e'.

{}From tables~\ref{tab:1} and \ref{tab:2}
it can be seen immediately that the result in the HV scheme respects
Fierz-symmetry. Namely, the entries for the Fierz-conjugated
diagrams (1a, 1b), (2a, 2b), (2c, 2d, 2e'), (2e, 2f), and (2g, 2g', 2g'')
are equal. In the case of the NDR scheme, we have the relations
\beqn
\Psi_{1b,\,2b,\,2d,\,2f}^{NDR} & = & (1+\ve)\,
\Psi_{1a,\,2a,\,2c,\,2e}^{NDR} \, , \\
\tvs
\Psi_{2e'}^{NDR} & = & \Psi_{2c}^{NDR} \,, \label{eq:3.11b} \\
\tvs
\Psi_{2g}^{NDR} & = & (1+\ve)\,\Psi_{2g'}^{NDR}
\, = \, (1+\ve)^2\,\Psi_{2g''}^{NDR} \, .
\label{eq:3.11a}
\eeqn

\subsection{Current-current operators in the NDR scheme}

As was already mentioned above, diagrams 2e and 2e' contain
traces with an odd
number of $\gamma_5$'s, and thus a direct evaluation in the NDR scheme
is not possible. We can, however, use a ``trick'' in order to
circumvent this problem. The trouble stems from the fact that $Q_1$
is in the colour non-singlet form. For 2--point functions of only
colour singlet operators the problematic diagrams 2e and 2e' do not arise.
Therefore, a solution lies in choosing the basis $(\wt Q_1,\,Q_2)$,
enabling one to calculate all diagrams without $\gamma_5$-problems
\cite{PD:91}. The price to pay is the fact that this
basis is no longer closed under renormalization, and we have to
explicitly add two evanescent operators $E_1\equiv\wt Q_1-Q_1$ and
$E_2\equiv\wt Q_2-Q_2$.

Now, the original and the new basis read
\begin{equation}
Q \; = \; (\,Q_1, \,Q_2, \,E_1, \,E_2) \qquad \hbox{and} \qquad
\wt Q \; = \; (\,\wt Q_1, \,Q_2, \,E_1, \,E_2) \,.
\label{eq:3.12}
\end{equation}
The 1--loop renormalization of these two bases can be obtained by
calculating matrix elements of $Q$ and $\wt Q$ between free quark
states. A straightforward computation leads to
$<\wt Q^B> \; = \; \wt M \wt Q^{\mbox{\rms Tree}}$ with
\begin{equation}
\wt M \; = \; 1 + \ap\,\biggl(\,\frac{3}{4\ve}-\frac{7}{4}\,\biggr)\!
\left( \begin{array}{cccc}
-1/N & 1 & 0 & 1 \\  1 & -1/N & -1 & 0 \\
0 & 0 & -1/N & 1 \\  0 & 0 & 1 & -1/N
\end{array} \right) \,.
\label{eq:3.13}
\end{equation}
The corresponding matrix $M$ for the $Q$-basis is the same except
for zeros in the entries (1,4) and (2,3). Thus, the finite
shift $\wt{\Delta r}$ of eq.~\eqn{eq:2.16}, mediating between the
$\wt Q$-basis and the $Q$-basis is given by
\begin{equation}
\wt{\Delta r} \; = \; \frac{7}{4}
\left( \begin{array}{cccc}
0 & 0 & 0 & -1 \\   0 & 0 & 1 & 0 \\
0 & 0 & 0 & 0 \\   0 & 0 & 0 & 0
\end{array} \right) \,.
\label{eq:3.14}
\end{equation}
For the 1--loop anomalous dimension matrices we find
\begin{equation}
\wt\gamma^{(1)} \; = \; 2\wt Z_1^{(1)} \; = \; \frac{3}{2}
\left( \begin{array}{cccc}
-1/N & 1 & 0 & 1 \\  1 & -1/N & -1 & 0 \\
0 & 0 & -1/N & 1 \\  0 & 0 & 1 & -1/N
\end{array} \right) \,.
\label{eq:3.15}
\end{equation}
and $\gamma^{(1)}$ is the same with the entries (1,4) and (2,3)
again being zero.

We are now in a position to calculate all matrices in the NDR scheme.
$A$ and $C$ are not scheme-dependent and therefore given by
eqs.~\eqn{eq:3.8} and \eqn{eq:3.9}, with all other entries being zero.
The matrices $B^{NDR}$, $\wt B^{NDR}$, and $\wt D^{NDR}$, can be
obtained from tables~\ref{tab:1} and \ref{tab:2}. The result is
\begin{equation}
B^{NDR} \; = \; -\,\frac{653}{60}\,A \,, \qquad
\wt D^{NDR} \; = \; \frac{1}{45}\,N C_f
\left( \begin{array}{cc}
6N & -175 \\ -175 & 6N
\end{array} \right) \,,
\label{eq:3.16}
\end{equation}

\begin{equation}
\wt B^{NDR} \; = \; -\;\frac{1}{675}\,N
\left( \begin{array}{cccc}
653N & 593  &  60N &  60  \\
593  & 653N & -60  & -60N \\
 60N & -60  & 120N & 120  \\
 60  & -60N & 120  & 120N
\end{array} \right) \,,
\label{eq:3.17}
\end{equation}
where for $B^{NDR}$ and $\wt D^{NDR}$ we only need the projection on
the physical subspace. Next, we calculate the combination
$P\equiv(\gamma^{(1)}B+B\,\gamma^{(1)^T})/2$ which appears in
eq.~\eqn{eq:2.8}:
\begin{equation}
P^{NDR} \; = \; -\,\frac{653}{225}\,N C_f
\left( \begin{array}{cc} 0 & 1 \\ 1 & 0 \end{array} \right) \,,
\qquad
\wt P^{NDR} \; = \; -\,\frac{593}{225}\,N C_f
\left( \begin{array}{cc} 0 & 1 \\ 1 & 0 \end{array} \right) \,.
\label{eq:3.18}
\end{equation}
In addition, the contribution from the shift $\wt{\Delta r}$ in
eq.~\eqn{eq:2.16}, $\wt{\Delta r} A+A\,\wt{\Delta r}^T$, vanishes.
Combining everything, we obtain for $D^{NDR}$,
\begin{equation}
D^{NDR} \; = \; \wt D^{NDR} - \wt P^{NDR} + P^{NDR} \; = \;
\frac{1}{45}\,N C_f
\left( \begin{array}{cc}
6N & -187 \\ -187 & 6N
\end{array} \right) \,.
\label{eq:3.19}
\end{equation}
Using this result together with eqs.~\eqn{eq:3.3} and \eqn{eq:3.4},
we can deduce the entries $D_{2e}^{NDR}$ and $D_{2e'}^{NDR}$ of
tab.~\ref{tab:2}, which were not calculable
directly. Finally, we have
\begin{equation}
X^{NDR} \; = \; \frac{2}{75}\,N C_f
\left( \begin{array}{cc}
5N & -47 \\ -47 & 5N
\end{array} \right) \,.
\label{eq:3.20}
\end{equation}

As a test for this result, we can use the relation \eqn{eq:2.16}
between the HV and the NDR scheme. The corresponding matrix
$\Delta r$ for this case can be obtained from eq. (3.9) of ref.
\cite{BJLW:92b} and eq. (4.25) of \cite{BJL:93}:
\begin{equation}
\Delta r \; = \; \frac{1}{2}
\left( \begin{array}{cc}
-1/N & 1 \\ 1 & -1/N
\end{array} \right) \,.
\label{eq:3.21}
\end{equation}
It turns out that the relation \eqn{eq:2.16} is indeed satisfied,
providing a strong check of our result.

\subsection{The diagonal basis}

Further insight in our results for the current-current operators
can be gained by transforming to the diagonal basis
$Q_{\pm}\equiv(Q_2\pm Q_1)/2$ \cite{ACMP:81,BW:90}. Following
ref.~\cite{BW:90}, we define the scheme-invariant operators
\begin{equation}
\overline Q_{\pm} \; \equiv \; \biggl[\,1+\ap\,B_{\pm}\,
\biggr]\,Q_{\pm} \,,
\label{eq:3.23}
\end{equation}
where the scheme-dependent coefficients $B_{\pm}$ are given by
\begin{equation}
B_{\pm}^{HV} \; = \;  \frac{7}{8}\,\biggl(\pm\,1-\frac{1}{N}\,\biggr) \,;
\qquad
B_{\pm}^{NDR}\; = \; \frac{11}{8}\,\biggl(\pm\,1-\frac{1}{N}\,\biggr) \,.
\label{eq:3.24}
\end{equation}
The 2--point functions in that basis are explicitly scheme-independent
and read
\begin{equation}
\overline\Psi_{\pm\pm} \; = \; \frac{1}{2}\,\biggl(\,1+\ap\,2B_{\pm}\,\biggr)
\biggl[\,\Psi_{11}\pm\Psi_{12}\,\biggr] \,,
\label{eq:3.25}
\end{equation}
if we take advantage of the result $\Psi_{2c}=\Psi_{2e'}$.

Using this expression together with eq.~\eqn{eq:2.10}, the spectral
functions
$\overline\Phi_{\pm\pm}\equiv {1\over\pi}\mbox{\rm Im}\overline\Psi_{\pm\pm}$
turn out to be
\begin{equation}
\overline\Phi_{\pm\pm}(s,\mu^2) \; = \;
\theta(s)\,\frac{s^4}{(4\pi)^6}\,A_{\pm}\biggl\{\,1+\ap\biggl[\,
\frac{3}{2}\biggl(\pm\,1-\frac{1}{N}\biggr)\ln\biggl\vert\frac{s}{\mu^2}
\biggr\vert+\frac{3}{4}N\mp\frac{101}{20}+\frac{43}{10}\frac{1}{N}\,
\biggr]\,\biggr\} \,,
\label{eq:3.26}
\end{equation}
with $A_{\pm}=2N(N\pm1)/45$.
The coefficient of the logarithm is, of course, just equal to the
leading-order anomalous dimensions of $Q_{\pm}$, $\gamma_{\pm}^{(1)}$.

The corresponding coefficient functions $C_{\pm}(\mu^2,M_W^2)$ have
been calculated in refs.~\cite{ACMP:81,BW:90}. We find it convenient to
split up the Wilson-coefficients in two factors, one solely depending
on $\mu^2$ and the other on $M_W^2$:
$C_{\pm}(\mu^2,M_W^2) \; = \; C_{\pm}(\mu^2)\,C_{\pm}(M_W^2)$.
The two factors are given by
\begin{equation}
C_{\pm}(\mu^2) \; = \; \alpha_s(\mu^2)^{\gamma^{(1)}_{\pm}/\beta_1}
\biggl[\,1-\frac{\alpha_s(\mu^2)}{4\pi}\,R_{\pm}\,\biggr] \,,
\quad
C_{\pm}(M_W^2) \; = \; \alpha_s(M_W^2)^{-\gamma^{(1)}_{\pm}/\beta_1}
\biggl[\,1+\frac{\alpha_s(M_W^2)}{4\pi}\,R_{\pm}\,\biggr] \,.
\label{eq:3.27}
\end{equation}
The NLO correction $R_{\pm}$ can be found in ref.~\cite{BW:90}.

Using this result, we are in a position to form the scale-independent
spectral functions of eq.~\eqn{eq:2.14}, $\wh\Phi_{\pm\pm}(s)=
C_{\pm}^2(M_W^2)\,C_{\pm}^2(s)\,\overline\Phi_{\pm\pm}(s)$:
\begin{equation}
\wh\Phi_{\pm\pm}(s) \; = \; \theta(s) \,
\frac{s^4}{(4\pi)^6}\,\alpha_s(s)^{2\gamma^{(1)}_{\pm}/\beta_1}
C_{\pm}^2(M_W^2)\biggl[\,A_{\pm}+\aps\,\wh X_{\pm}\,\biggr] \,.
\label{eq:3.28}
\end{equation}
Setting $f=3$, we find for the NLO contributions $\wh X_{\pm}$:
\begin{eqnarray}
\wh X_{+} & = & \frac{C_f}{\beta_1^2}\,\biggl[\,\frac{121}{540}N^4-
\frac{30917}{16200}N^3+\frac{11173}{5400}N^2-\frac{781}{600}N+\frac{5}{12}
\,\biggr] \,, \label{eq:3.29} \\
\smvs
\wh X_{-} & = & \frac{C_f}{\beta_1^2}\,\biggl[\,\frac{121}{540}N^4+
\frac{22997}{16200}N^3-\frac{8143}{5400}N^2+\frac{641}{600}N-\frac{5}{12}
\,\biggr] \,.
\label{eq:3.30}
\end{eqnarray}
For $N=3$ the two spectral functions simplify to
\begin{eqnarray}
\wh \Phi_{++}(s) & = & \theta(s)\,
\frac{8}{15}\,\frac{s^4}{(4\pi)^6}\,\alpha_s(s)^{-4/9}\,
C_{+}^2(M_W^2) \biggl[\,1-\frac{3649}{1620}\,\aps\,\biggr] \,,
\label{eq:3.31} \\
\smvs
\wh \Phi_{--}(s) & = & \theta(s)\,
\frac{4}{15}\,\frac{s^4}{(4\pi)^6}\,\alpha_s(s)^{8/9\phantom{-}}\,
C_{-}^2(M_W^2) \biggl[\,1+\frac{9139}{810}\,\aps\,\biggr] \,.
\label{eq:3.32}
\end{eqnarray}

Let us comment briefly on the implications of our results.
Again taking $\alpha_s(s)/\pi\approx 0.1$, at the NLO we find a moderate
suppression of $\wh\Phi_{++}$ by roughly 20\%, whereas $\wh\Phi_{--}$
acquires a huge enhancement on the order of 100\%, including the
coefficient functions at $M_W$, $C_{\pm}(M_W^2)$, which only have a
minor effect. Because $\wh\Phi_{++}$ solely receives contributions
from $\Delta I=3/2$, and $\wh\Phi_{--}$ is a mixture of both
$\Delta I=1/2$ and $\Delta I=3/2$, this pattern of the radiative
corrections entails a strong enhancement of the $\Delta I=1/2$
amplitude. Hence, we are provided with a very promising picture
for the emergence of the $\Delta I=1/2$--rule.

Analyzing our result from the point of view of the large-$N$
expansion~\cite{BBG:87,BBG:87a},
we see that at leading-order, the corrections to $\wh\Phi_{++}$ and
$\wh\Phi_{--}$ are equal \cite{PI:89,PD:91}, meaning that the
$\Delta I=1/2$--rule is missed completely. This situation is partly
remedied if the large next-to-leading corrections in $1/N$ are
taken into account.

\newsection{Full result including penguins}

To obtain the complete result, we have to add to our basis the
following four penguin operators
\begin{eqnarray}
Q_{3} & = & \left( \bar s d \right)_{\rm V-A}
   \sum_{q} \left( \bar q q \right)_{\rm V-A} \,,
\qquad
Q_{4} \;= \;\left( \bar s_{\alpha} d_{\beta}  \right)_{\rm V-A}
   \sum_{q} \left( \bar q_{\beta}  q_{\alpha} \right)_{\rm V-A} \,, \nn \\
Q_{5} & = & \left( \bar s d \right)_{\rm V-A}
   \sum_{q} \left( \bar q q \right)_{\rm V+A} \,,
\qquad
Q_{6} \;= \;\left( \bar s_{\alpha} d_{\beta}  \right)_{\rm V-A}
   \sum_{q} \left( \bar q_{\beta}  q_{\alpha} \right)_{\rm V+A} \,,
\label{eq:4.1}
\end{eqnarray}
which arise from the current-current operators in the process of
renormalization. The corresponding Fierz-transformed operators,
again being needed for the calculation in the NDR scheme can be
found in ref.~\cite{BJLW:92b}.

The 2--point functions for $(V-A)\otimes(V-A)$ operators including
$Q_3$ and $Q_4$ are given by
\begin{eqnarray}
\Psi_{13} & = & N^2 \Psi_{1b}+N^2 C_f\Big[\,4\Psi_{2b}+2\Psi_{2d}\,
\Big] \,, \label{eq:4.2} \\
\Psi_{14} & = & N \Psi_{1b}+N C_f\Big[\,4\Psi_{2b}+4\Psi_{2d}+
2\Psi_{2f}\,\Big] \,, \label{eq:4.3} \\
\Psi_{23} & = & \Psi_{14} + 2N C_f \Psi_{2g} \,, \qquad
\Psi_{24} \;= \;\Psi_{13} + fN C_f \Psi_{2g'} \,, \label{eq:4.4} \\
\Psi_{33} & = & f \Psi_{11} + 2 \Psi_{23} \,, \qquad
\Psi_{34} \;= \;f \Psi_{12} + 2 \Psi_{24} \,, \label{eq:4.5} \\
\Psi_{44} & = & f \Psi_{11} + 2 \Psi_{14} + f^2 N C_f \Psi_{2g''} \,.
\label{eq:4.6}
\end{eqnarray}
In the expression for $\Psi_{33}$, we have used the relation
$\Psi_{2c}=\Psi_{2e'}$.

The 2--point functions with insertions of the $(V-A)\otimes(V+A)$
operators $Q_5$ and $Q_6$ can be calculated analogously. For simplicity,
we don't give their contributions in detail, but just state the final
results. In the HV scheme, we find:
\begin{equation}
A \; = \; \frac{4}{45}\,N
\left( \begin{array}{cccccc}
N & 1 & N & 1 & 0 & 0 \\
1 & N & 1 & N & 0 & 0 \\
N & 1 & f N + 2 & 2 N + f & 0 & 0 \\
1 & N & 2 N + f & f N + 2 & 0 & 0 \\
0 & 0 & 0 & 0 & f N & f \\
0 & 0 & 0 & 0 & f & f N
\end{array} \right) \,,
\label{eq:4.8}
\end{equation}

\begin{equation}
B^{HV} \; = \; -\,N
\left( \begin{array}{cccccc}
\frac{1637N}{1575} & \frac{1637}{1575} & \frac{1637N}{1575} &
\frac{1637}{1575} & \frac{-2N}{315} & \frac{-2}{315} \\
\tvs
\frac{1637}{1575} & \frac{1637N}{1575} & \frac{1637}{1575} &
\frac{1637N}{1575} & \frac{-2}{315} & \frac{-2N}{315} \\
\tvs
\frac{1637N}{1575} & \frac{1637}{1575} & \frac{1637(fN+2)}{1575} &
\frac{1637(2N+f)}{1575} & \frac{-2(fN+2)}{315} & \frac{-2(2N+f)}{315} \\
\tvs
\frac{1637}{1575} & \frac{1637N}{1575} & \frac{1637(2N+f)}{1575} &
\frac{1637(fN+2)}{1575} & \frac{-2(2N+f)}{315} & \frac{-2(fN+2)}{315} \\
\tvs
\frac{-2N}{315} & \frac{-2}{315} & \frac{-2(fN+2)}{315} &
\frac{-2(2N+f)}{315} & \frac{1637fN}{1575}-\frac{4}{315} &
\frac{1637f}{1575}-\frac{4N}{315} \\
\tvs
\frac{-2}{315} & \frac{-2N}{315} & \frac{-2(2N+f)}{315} &
\frac{-2(fN+2)}{315} & \frac{1637f}{1575}-\frac{4N}{315} &
\frac{1637fN}{1575}-\frac{4}{315}
\end{array} \right) \,.
\label{eq:4.9}
\end{equation}
The matrix $C$ satisfies the relation \eqn{eq:2.9} with $\gamma^{(1)}$
given in ref.~\cite{BJLW:92b}, and we don't list it explicitly. The
matrix $D^{HV}$ has been relegated to the appendix. Inserting these into
eq.~\eqn{eq:2.8}, we obtain for $X^{HV}$:
\begin{equation}
X^{HV} = N C_f \!
\left( \begin{array}{cccccc}
\frac{2N}{15} & \frac{-242}{225} & \frac{2N}{15} & \frac{-242}{225} &0&0 \\
\tvs
\frac{-242}{225} & \frac{2N}{15}-\frac{242}{2025} & \frac{-2662}{2025} &
\frac{2N}{15}-\frac{242f}{2025} & 0 & \frac{-242f}{2025} \\
\tvs
\frac{2N}{15} & \frac{-2662}{2025} & \frac{2fN}{15}-\frac{5324}{2025} &
\frac{4N}{15}-\frac{2662f}{2025} & 0 & \frac{-484f}{2025} \\
\tvs
\frac{-242}{225} & \frac{2N}{15}-\frac{242f}{2025} & \frac{4N}{15}-
\frac{2662f}{2025} & \frac{2fN}{15}-\frac{242f^2}{2025}-\frac{484}{225} &
0 & \frac{-242f^2}{2025} \\
\tvs
0 & 0 & 0 & 0 & \frac{2fN}{15} & \frac{38f}{25} \\
\tvs
0 & \frac{-242f}{2025} & \frac{-484f}{2025} & \frac{-242f^2}{2025} &
\frac{38f}{25} & \frac{322fN}{225}-\frac{242f^2}{2025}
\end{array} \right) .
\label{eq:4.10}
\end{equation}

The treatment to work around the $\gamma_5$ problem in the NDR scheme
for the full basis parallels the method used in the current-current
case. We can choose the basis used in ref.~\cite{PD:91},
$(\,\wt Q_1,\,Q_2,\,Q_3,\,\wt Q_4,\,Q_5,\,\wt Q_6)$, which does not contain
colour non-singlet operators, thus not posing $\gamma_5$ problems for the
computation of the 2--point function.
However, because this basis does not close under
renormalization, this time, for each of the six operators we have to
add to the basis an evanescent operator, implying that at intermediate
steps of the calculation we have to work with $12\times 12$ matrices.
We skip the unilluminating details of this computation and just present
our results.

As already stated, the matrices $A$ and $C$ do not depend on the scheme,
and agree with the $HV$ case. For $B^{NDR}$ we find
\begin{equation}
B^{NDR} \; = \; -\,N
\left( \begin{array}{cccccc}
\frac{653N}{675} & \frac{653}{675} & \frac{593N}{675} &
\frac{593}{675} & 0 & 0 \\
\tvs
\frac{653}{675} & \frac{653N}{675} & \frac{593}{675} &
\frac{593N}{675} & 0 & 0 \\
\tvs
\frac{593N}{675} & \frac{593}{675} & \frac{653fN+1186}{675} &
\frac{1186N+653f}{675} & 0 & 0 \\
\tvs
\frac{593}{675} & \frac{593N}{675} & \frac{1186N+653f}{675} &
\frac{653fN+1186}{675} & 0 & 0 \\
\tvs
0 & 0 & 0 & 0 & \frac{653fN}{675} & \frac{653f}{675} \\
\tvs
0 & 0 & 0 & 0 & \frac{653f}{675} & \frac{653fN}{675} \\
\end{array} \right) \,,
\label{eq:4.11}
\end{equation}
and $D^{NDR}$ can again be found in the appendix. From these
results we deduce for $X^{NDR}$:
\begin{equation}
X^{NDR} = N C_f \!
\left( \begin{array}{cccccc}
\frac{2N}{15} & \frac{-94}{75} & \frac{2N}{15} & \frac{-94}{75} &0&0 \\
\tvs
\frac{-94}{75} & \frac{2N}{15}-\frac{182}{2025} & \frac{-2902}{2025} &
\frac{2N}{15}-\frac{212f}{2025} & 0 & \frac{-212f}{2025} \\
\tvs
\frac{2N}{15} & \frac{-2902}{2025} & \frac{2fN}{15}-\frac{5804}{2025} &
\frac{4N}{15}-\frac{2962f}{2025} & 0 & \frac{-424f}{2025} \\
\tvs
\frac{-94}{75} & \frac{2N}{15}-\frac{212f}{2025} & \frac{4N}{15}-
\frac{2962f}{2025} & \frac{2fN}{15}-\frac{242f^2}{2025}-\frac{188}{75} &
0 & \frac{-242f^2}{2025} \\
\tvs
0 & 0 & 0 & 0 & \frac{2fN}{15} & \frac{74f}{75} \\
\tvs
0 & \frac{-212f}{2025} & \frac{-424f}{2025} & \frac{-242f^2}{2025} &
\frac{74f}{75} & \frac{94fN}{75}-\frac{242f^2}{2025}
\end{array} \right) .
\label{eq:4.12}
\end{equation}

The expressions for $X^{HV}$ and $X^{NDR}$ given above again do satisfy
the relation \eqn{eq:2.16} with $\Delta r$ taken from ref.~\cite{BJLW:92b}
and eq. (4.25) of \cite{BJL:93} to be
\begin{equation}
\Delta r \; = \; \frac{1}{2}
\left( \begin{array}{cccccc}
-\frac{1}{N} & 1 & 0 & 0 & 0 & 0 \\
\tvs
1 & -\frac{1}{N} & \frac{1}{6N} & -\frac{1}{6} & \frac{1}{6N} & -\frac{1}{6} \\
\tvs
0 & 0 & -\frac{2}{3N} & \frac{2}{3} & \frac{1}{3N} & -\frac{1}{3} \\
\tvs
0 & 0 & 1 & -\frac{1}{N} & 0 & 0 \\
\tvs
0 & 0 & 0 & 0 & -\frac{3}{N} & 3 \\
\tvs
0 & 0 & 0 & 0 & 2 & N-\frac{3}{N}
\end{array} \right) .
\label{eq:4.13}
\end{equation}
This test of our results provides us with great confidence as to their
correctness.

Let us make a few observations on the results thus obtained:
\begin{itemize}
\item Comparing the NLO matrices $X^{NDR}$ and $X^{HV}$ to the LO matrix
 $A$, we find huge corrections on the order of 200\% for the entries (1,2),
 (1,4), (2,3), (5,6), and (6,6). The difference between the NDR and HV
 scheme is small with respect to the large absolute value of the corrections.
 All other entries have moderate corrections $\lsim$~50\%.
\item If the number of flavours $f=3$, in the HV scheme we have the operator
 relation $Q_4=Q_3+Q_2-Q_1$. This leads to the following relation for the
 two point functions:
 $\Psi_{4i}^{HV}=\Psi_{3i}^{HV}+\Psi_{2i}^{HV}-\Psi_{1i}^{HV}$. The relation
 is satisfied by our result, providing an additional test. In the NDR scheme,
 the relation is broken by $\cal O(\alpha_s)$ corrections.\footnote{See also
 sect.~4.5 of ref.~\cite{BJL:93}}
\item Due to the factorization property of the $Q_6$ operator in the large-$N$
 limit, in this limit $\Psi_{66}$ can be related to a convolution of
 two 2--point functions for scalar currents \cite{PD:91}. This relation
 holds for our result but we shall return to this point in sect.~7.
\end{itemize}

\newsection{Numerical Results}

In this section, we shall provide the reader with a brief discussion of
the numerical implications of our results, and postpone a thorough
phenomenological analysis to a forthcoming publication.

Following the notation of refs.~\cite{BJLW:92a,BJL:93}, the
Wilson-coefficient functions for $\Delta S=1$ weak processes can
be decomposed as $C(s) = z(s) + \tau\,y(s)$, where
$\tau\equiv - \left(V_{td}^{\phantom{*}} V_{ts}^*\right)/
\left(V_{ud}^{\phantom{*}} V_{us}^*\right)$.
The coefficient function $z(s)$
governs the real part of the effective Hamiltonian, and $y(s)$,
parametrizes the imaginary part and governs e.g. the measure for
direct CP-violation in the $K$-system, $\ve'/\ve$.
We thus have two different quantities with the help of which we can
form the scale- and scheme-invariant combination $\wh\Phi(s)$ of
eq.~\eqn{eq:2.14}. Let us denote these two functions by:
\begin{equation}
\wh\Phi_z(s) \; = \; z^T(s)\,\Phi(s)\,z(s) \,; \qquad
\wh\Phi_y(s) \; = \; y^T(s)\,\Phi(s)\,y(s) \,.
\label{eq:5.1}
\end{equation}

For the numerical analysis, we shall consider the range
$Q\equiv\sqrt{s}=1-3\,\mbox{\rm GeV}$,
appearing as a natural scale for a QCD sum rule analysis of the $K$-system
\cite{PDPPR:91}. In this range, the coefficient functions have the
following structure:
\begin{itemize}
\item Above the charm threshold $m_c$, $z_3-z_6$ vanish, and $\wh\Phi_z$
 is only given as a product of $(z_1,\,z_2)$ and the current-current part
 of $\Phi(s)$. Below $m_c$, penguins are generated from the operator
 mixing, but the coefficient $z_3-z_6$ still remain small above
 $1\,\mbox{\rm GeV}$,
 such that $\wh\Phi_z(s)$ in the whole range is dominated by
 current-current operators.
\item In the case of $y(s)$, only $y_3-y_6$ are non-vanishing in the
 whole range considered. The coefficient for the $Q_6$ operator $y_6$
 dominates, but the other penguin operators also give noticeable contributions.
\end{itemize}

Since in this work we are mainly interested in the size of the radiative
corrections to the effective Hamiltonian, we write $\wh\Phi(s)$ as
\begin{equation}
\wh\Phi_{z,\,y}(s)\;=\;\wh\Phi^{(0)}_{z,\,y}(s)+\wh\Phi^{(1)}_{z,\,y}(s) \,,
\label{eq:5.2}
\end{equation}
where the superscripts $(0)$ and $(1)$ refer to the leading as well
as next-to-leading order respectively.
In fig.~3, we plot the ratios $\wh\Phi^{(1)}_{z}/\wh\Phi^{(0)}_{z}$ and
$\wh\Phi^{(1)}_{y}/\wh\Phi^{(0)}_{y}$ for $\La=200,\,300$, and $400$ MeV.
For simplicity, in fig.~3, we have not included the quark threshold at
$m_c$, but we work in a theory with $f=3$ up to $3\,\mbox{\rm GeV}$.
We have checked
that these threshold effects only cause a small change in the NLO
correction. Of course, as expected, the values for $\wh\Phi_{z,\,y}$
in the HV and NDR scheme exactly agree.

{}From fig.~3, we can see that in the region $Q=1-3\,\mbox{\rm GeV}$, and for a
central value $\La=300\,\mbox{\rm MeV}$, the radiative
QCD correction to $\wh\Phi_z$ ranges approximately between 40\% and 120\%,
whereas in the case of $\wh\Phi_y$ we find a correction on the order of
100\%--240\%. Because the 2--point function is constructed as the square
of the effective Hamiltonian, the actual corrections to
${\cal H}^{\Delta S=1}_{\mbox{\rms eff}}$
are only about half the corrections to the 2--point function. Therefore,
the perturbative QCD correction to the real part of the effective Hamiltonian
turns out to be 20\%--60\%, and for the imaginary part 50\%--120\%.

In sect.~3.3, we have demonstrated that the large $\al$ corrections correspond
to the $\Delta I=1/2$ part of the effective weak Hamiltonian\footnote{
This has also been extensively discussed in ref. \protect\cite{PD:91}.}.
The corrections to the $\Delta I=3/2$ part are identical to the
ones in the $\Delta S=2$ correlator
(both operators are in the same representation of the chiral group),
which, as shown in sect.~3.3 and the next section, are quite moderate
and negative. This implies that for the $\Delta I=1/2$--rule in
$K\rightarrow\pi\pi$ decays, we receive an additional large and
positive contribution, bringing theoretical calculations closer
to the experimental value.

The calculation of the imaginary part of $\cal H_{\mbox{\rms eff}}$,
no longer retains perturbative character, because of the large
corrections. Nevertheless, this does {\em not} completely spoil
existing determinations of weak matrix elements in the framework
of the $1/N$ expansion or chiral perturbation theory, for there, to a given
order in $1/N$ or the chiral expansion, the largest corrections
in $\alpha_s$ are completely summed to all orders.

\newsection{The $\ds$ operator}

For the case of $\ds$ transitions, things are somewhat simpler because
there is only one operator. We take this operator to be
\begin{equation}
Q_{\ds} \; \equiv \; \frac{1}{2}\,\Big[\,
\left( \bar s d \right)_{\rm V-A}
\left( \bar s d \right)_{\rm V-A} +
\left( \bar s_{\alpha} d_{\beta}  \right)_{\rm V-A}
\left( \bar s_{\beta}  d_{\alpha} \right)_{\rm V-A}\,\Big] \,.
\label{eq:6.1}
\end{equation}
This definition might seem unfamiliar, but we shall discuss in the
following, why it is convenient. First, let us note that
it renormalizes into itself, even in a general dimension\footnote{Apart
from evanescent terms which are taken care of by the projection discussed
in sect.~3.1}, and it is obviously Fierz-symmetric. Apart from the
quark content it has the same structure as $Q_{+}$.

Collecting the contributing diagrams, and making use of the relation
\eqn{eq:3.11b}, the 2--point function of $Q_{\ds}$ is given by
\begin{equation}
\Psi_{\ds} \; = \; \Psi_{11} + \Psi_{12} + \Psi_{13} + \Psi_{14} \,.
\label{eq:6.2}
\end{equation}
{}From this expression, we obtain the following values for $A$, $B$, and $X$:
\begin{eqnarray}
A_{\ds} & = & \frac{8}{45}\,N\Big(N+1\Big) \,, \label{eq:6.3} \\
\smvs
B_{\ds}^{HV} & = & -\,\frac{3274}{1575}\,N\Big(N+1\Big) \,;
\qquad
B_{\ds}^{NDR} \; = \; -\,\frac{1246}{675}\,N\Big(N+1\Big) \,,\label{eq:6.4} \\
\smvs
X_{\ds}^{HV} & = & N C_f\,\biggl(\frac{4N}{15}-\frac{484}{225}\biggr) \,;
\qquad
X_{\ds}^{NDR} \; = \; N C_f\,\biggl(\frac{4N}{15}-\frac{188}{75}\biggr) \,.
\label{eq:6.5}
\end{eqnarray}
As was already remarked at the end of the last section, up to a
multiplicity factor 4, these quantities agree with the corresponding
expressions for $\Psi_{++}$ of sect.~3.3, if we would refrain from
performing the rotation of eq.~\eqn{eq:3.23} to a scheme-invariant basis.
$C_{\ds}$ again respects eq.~\eqn{eq:2.9}, and the relation between
schemes \eqn{eq:2.16} is also satisfied with
\begin{equation}
\Delta r_{\ds} \; = \; \frac{1}{2}\,\biggl(\,1-\frac{1}{N}\,\biggr) \,,
\label{eq:6.6}
\end{equation}
being easily obtained from ref.~\cite{BJLW:92b}.

In the HV scheme, we could have worked with the operator
\begin{equation}
O_{\ds} \; \equiv \; \left( \bar s d \right)_{\rm V-A}
                     \left( \bar s d \right)_{\rm V-A} \,,
\label{eq:6.7}
\end{equation}
in 4 dimensions being equivalent to $Q_{\ds}$, since in HV Fierz-symmetry
is respected for current-current operators, and $O_{\ds}$ renormalizes into
itself. This is not true for the NDR scheme and we would have to go through
a similar procedure as described in sect.~3.2. Namely, augmenting the basis
by an evanescent operator $E_{\ds}=\wt O_{\ds}-O_{\ds}$ and including this
operator for renormalization, because it induces additional contributions
to the physical subspace. We have checked that this treatment leads to the
same value for $X^{NDR}_{\ds}$. Let us point out that this only concerns
the quantities $B^{NDR}_{\ds}$ and $D^{NDR}_{\ds}$. The anomalous dimensions
of $Q_{\ds}$ and $O_{\ds}$ in the NDR scheme agree even at NLO.

In order to be able to form the scheme-independent combination
$\wh\Phi_{\ds}(s)$, we need the Wilson-coefficient function for $\ds$
processes at NLO. It can be obtained from refs.~\cite{BJW:90,HN:93} for
internal top and charm quark exchange in the box-diagram. The mixed
charm-top contribution is not yet available at NLO. However, here we
shall not pursue this any further, but use the strategy of ref.~\cite{BJW:90}
of defining a scale- and scheme-invariant operator for $\ds$. This operator
is given by
\begin{equation}
\wh Q_{\ds} \; \equiv \; \alpha_s(\mu^2)^{\gamma^{(1)}_{\ds}/\beta_1}
\biggl[\,1-\frac{\alpha_s}{4\pi}\,Z\,\biggr]\,Q_{\ds} \,,
\label{eq:6.8}
\end{equation}
where $\gamma^{(1)}_{\ds}=\gamma^{(1)}_{+}$ is the LO anomalous dimension
of $Q_{\ds}$. The finite NLO correction $Z$ depends on the scheme, and
can be found in \cite{BJW:90}. The matrix element of this operator is
directly parametrized in terms of the {\em scheme-invariant} $B$-parameter
$B_K$ for $K^0-\overline{K^0}$-mixing.

Calculating the spectral function $\wh\Phi_{\ds}(s)$ for $\wh Q_{\ds}$,
we obtain
\begin{equation}
\wh\Phi_{\ds}(s) \; = \; \theta(s) \,
\frac{s^4}{(4\pi)^6}\,\alpha_s(s)^{2\gamma^{(1)}_{\ds}/\beta_1}
\biggl[\,A_{\ds}+\aps\,\wh X_{\ds}\,\biggr] \,,
\label{eq:6.9}
\end{equation}
with $\wh X_{\ds}=4\,\wh X_{+}$, and $\wh X_{+}$ being given in
eq.~\eqn{eq:3.29}. This function is explicitly scheme-invariant,
and for $N=3$ takes the form
\begin{equation}
\wh \Phi_{\ds}(s) \; = \; \theta(s)\,
\frac{32}{15}\,\frac{s^4}{(4\pi)^6}\,
\alpha_s(s)^{-4/9} \biggl[\,1-\frac{3649}{1620}\,\aps\,\biggr] \,.
\label{eq:6.11}
\end{equation}
Because both, $Q_{+}$ and $Q_{\ds}$, have the same chiral representation,
as expected, apart from a global factor, their spectral functions
agree. We observe that the NLO QCD-correction is negative and on the
order of 20\%, for $\als/\pi\approx 0.1$.

\newsection{Discussion}

Our work improves and completes the 2--point function evaluation
of ref.~\cite{PD:91} with two major additions:
the recently calculated NLO corrections to the Wilson-coefficient functions
have been taken into account and, moreover, we have incorporated the
missing contributions from evanescent operators.
The final results are then renormalization scheme- and scale-independent
at the NLO, and, therefore, constitute the first complete calculation
of weak non-leptonic observables at the NLO, without any hadronic
ambiguity.

It is worthwhile to make a comparison with the results of ref.~\cite{PD:91}.
In this work, the NDR scheme was used and the calculation of the 2--point
function for the $\Delta S=1$ case was performed in the basis
\begin{equation}
\wt Q \; = \; \Big(\,\wt Q_1,\,Q_2,\,Q_3,\,\wt Q_4,\,Q_5,\,\wt Q_6\Big) \,,
\label{eq:7.1}
\end{equation}
not being plagued by problems with $\gamma_5$.
The topologies present in that calculation were
1a, 1b, 2a, 2b, 2c, 2d, 2f, and 2g.  We have reproduced the results
for all these  diagrams and we fully agree with ref.~\cite{PD:91}.
However, as already remarked in sects.~3 and 4, in the NDR scheme
the basis \eqn{eq:7.1} does not close under renormalization,
and we had to add contributions from evanescent operators which were not
included in \cite{PD:91}.
In the notation of sect.~3, these evanescent contributions shift the
matrices $\wt B^{NDR}$ and $\wt D^{NDR}$ to $B^{NDR}$ and $D^{NDR}$,
and therefore the final correction $X^{NDR}$ gets changed.
In addition, the relation
$\wt Q_4=Q_3+Q_2-\wt Q_1$ for $f=3$ was used in  ref.~\cite{PD:91}
 to eliminate $\wt Q_4$.
As it stands this relation is valid on the operator level.

Removing the entries for $Q_4$ from our matrices $B^{NDR}$, $D^{NDR}$, and
$X^{NDR}$, and setting $f=3$, one can easily see the differences with
the results of ref.~\cite{PD:91}.
The evanescent contributions have changed
the entries (1,2), (1,3), and (6,6) of $B^{NDR}$,
and (1,2) and (6,6) in $D^{NDR}$.
The differences in $B^{NDR}$ propagate via the matrix
 products of eq.~\eqn{eq:2.8} into most entries of $X^{NDR}$:
all entries except for (5,5) and
 the trivial zeros in (1,5), (2,5), and (3,5) are different.
The final numerical differences are however not big, since the
most sizeable $\al$ corrections were already included in the original
calculation of ref.~\cite{PD:91}.

In the large-$N$ limit the operator $Q_6$ factorizes in the product of
two current operators. Therefore, the 2--point function $\Psi_{66}$ can
be calculated as a convolution of two current-correlators.\footnote{For
details see ref.~\cite{PD:91}} This was used in \cite{PD:91} as a check
for $\Psi_{66}$ at the leading order in $1/N$.
In fact, the large-$N$ limit result of ref.~\cite{PD:91} was
already a full NLO calculation, since the corresponding
anomalous dimension $\gamma_{66}$ was already known at the NLO
(it is related to the quark-mass anomalous dimension in the large-$N$
limit) and it was correctly taken into account. Although our results for
$B_{66}^{NDR}$ and $D_{66}^{NDR}$ show a discrepancy to the result of
\cite{PD:91} even at the leading order in $1/N$, the combination
$X_{66}^{NDR}$ only deviates from the result of \cite{PD:91} by
subleading terms in the $1/N$-expansion, hence fulfilling the test
in both cases.
The differences at intermediate steps of the calculation
($B_{66}^{NDR}$ and $D_{66}^{NDR}$) stem from the fact that
a different form of the operator ($Q_6$ or $\wt Q_6$) is being used,
but the final physical result is of course identical.

For the $\ds$ operator the evanescent contributions result in changes in
all next-to-leading quantities $B_{\ds}^{NDR}$, $D_{\ds}^{NDR}$, and
$X_{\ds}^{NDR}$. The coefficient of the NLO correction to
$\overline \Phi_{\ds}$, eq.~\eqn{eq:6.11}, was found to be $-1217/810$
in ref.~\cite{PD:91}, compared to $-3649/1620$ in our case. This
correction was used in ref.~\cite{PDPPR:91} for a sum rule determination
of $B_K$. The effect of our new result would be to slightly further
reduce the value of $B_K$ obtained in ref.~\cite{PDPPR:91}.

Qualitatively, the conclusions of ref.~\cite{PD:91} remain
unchanged. At the NLO, the $\Delta I=1/2$ piece of the
$\Delta S=1$ effective Hamiltonian gets a huge positive correction,
while the gluonic effects in the $\Delta I=3/2$ (and $\Delta S=2$)
operator are moderate and negative. Together with the previously known
enhancement of the Wilson-coefficient \cite{ACMP:81,BJLW:92a,BW:90,BJL:93},
this provides a very suggestive explanation of the observed enhancement of
$\Delta I=1/2$ transitions in $K$ decays.

As explicitly shown in fig.~3, the NLO gluonic contributions
are even more important in the CP-violating piece of the weak
$\Delta S=1$ Hamiltonian; the reason being that this part is dominated
by the penguin operator $Q_6$, which gets the largest correction.
As shown in ref.~\cite{PD:91}, this enhancement is further
reinforced\footnote{Thanks to the factorization property of the
penguin operator in the large-$N$ limit, the $\cO(\al^2)$
correction is also known in this limit.}
at $\cO(\al^2)$, indicating a blow-up of the
perturbative series in this case.
Fortunately, this non-perturbative character does not completely prevent
the feasibility of a reliable determination of CP-violating effects,
since these leading contributions in $1/N$ can be resummed to all
orders.

In a series of articles \cite{Ste:87,DJS:89,NS:89,NS:91} a more
phenomenological description of the $\Delta I=1/2$--rule was advocated.
The key idea is to rewrite the 4--quark operators as a product of
diquark-anti-diquark operators by means of Fierz-transformations, and
treating the diquark as an effective particle, similar to the
constituent quarks. The important observation then lies in the dominance
of pseudoscalar diquark matrix elements for low momentum transfer over
axialvector meson matrix elements which are proportional to the momentum.
All low energy decays in which diquarks can participate show the
enhancement of $\Delta I=1/2$ amplitudes and a surprisingly good
description of those processes was obtained.

An inspection of our results at the diagrammatic level reveals the
following pattern for the origin of large corrections: all 2--point
functions (except those which vanish at lowest order)
receive contributions from the self-energy diagrams 2a or 2b,
as well as from the quark-antiquark vertex corrections 2c, 2d, or 2e'.
These contributions cancel to a fair amount:
\begin{eqnarray}
C_{2\cdot 2a+2c} \; = \; C_{2\cdot 2a+2e'} \; = \;
C_{2\cdot 2b+2d} & = & 0 \,, \\
\smvs
D^{NDR}_{2\cdot 2a+2c} \; = \; D^{NDR}_{2\cdot 2a+2e'} \; = \;
D^{NDR}_{2\cdot 2b+2d} & = & \frac{1}{15} \,, \\
\smvs
D^{HV}_{2\cdot 2a+2c} \; = \; D^{HV}_{2\cdot 2a+2e'} \; = \;
D^{HV}_{2\cdot 2b+2d} & = & \frac{7}{45} \,.
\end{eqnarray}
The quark-quark and
antiquark-antiquark correlation diagrams 2e or 2f already by themselves
are the biggest terms, and due to the partial cancellation of self-energy
and current-vertex diagrams, we receive large corrections wherever
quark-quark correlations can contribute. Note that these diagrams are
subleading in $1/N$. The penguin diagrams 2g, 2g', and 2g'' generally
have small impact on the 2-point functions.

This structure of the radiative corrections to 2--point functions of
$\Delta S=1$ and $\ds$ operators allows for a deeper understanding,
why the description of non-leptonic weak decays in terms of diquarks
was so successful as far as the $\Delta I=1/2$--rule is concerned.
In this framework, by working with effective diquarks, the quark-quark
correlations were phenomenologically
summed up to all orders in the strong coupling.
Since these are the dominant corrections to the 2--point functions,
summing them up provides us with a very good physical picture of the
underlying QCD dynamics. As such, though, the statement in question,
like the diquark-current itself,\footnote{See the related discussion in
ref.~\cite{DJS:89}.} is gauge-dependent. In fact, the gauge-invariant
combinations of diagrams are $(2\cdot 2a+2c)$, $(2\cdot 2a+2e')$,
$(2e+2e')$, as well as their corresponding Fierz-conjugates. However,
now the gauge-independent combination involving the quark-quark
correlations dominates the other terms even more drastically
(by one order of magnitude):
\begin{eqnarray}
&& C_{2e+2e'} \; = \; C_{2d+2f} \; = \;  \frac{2}{15} \,, \\
\smvs
&& D^{NDR}_{2e+2e'} \; = \; -\,\frac{193}{90} \,, \qquad
D^{NDR}_{2d+2f} \; = \;  -\,\frac{181}{90} \,, \\
\smvs
&& D^{HV}_{2e+2e'} \; = \; D^{HV}_{2d+2f} \; = \;  -\,\frac{1363}{630} \,.
\end{eqnarray}

A full QCD calculation has been possible because of the inclusive character
of the defined 2--point functions.
Although only qualitative conclusions can be directly
extracted from these results, they are certainly important since they
rigorously point to the QCD origin of the infamous $\Delta I=1/2$--rule,
and, moreover, provide valuable information on the relative importance
of the different operators, which can be very helpful to attempt
more pragmatic calculations.
Obviously, a direct application of our results would be the use
of dispersion relations to extract ``more exclusive'' information from
the 2--point functions, following the methods developed
in refs. \cite{PD:85,GPD:85,PGD:86,PD:87,PI:89,PDPPR:91,PD:91}.
We plan to investigate this and other possible phenomenological
applications in the future.

\vskip 1cm \noindent
{\Large\bf Acknowledgement}

\noindent
We would like to thank A. J. Buras and E. de Rafael for discussions
and for reading the manuscript. M. J. would like to thank M. Neubert
and P. H. Weisz for discussion. The Feynman diagrams were drawn with
the aid of the program {\em feynd}, written by S. Herrlich. The work
of A.P. has been supported in part by CICYT (Spain), under grant
No. AEN93-0234.

\vskip 1cm \noindent
\appendix{\LARGE\bf\noindent Appendix}

\begin{displaymath}
D^{NDR} = N C_f \!
\left( \begin{array}{cccccc}
\frac{2N}{15} & \frac{-187}{45} & \frac{2N}{15} & \frac{-35}{9} &0&0 \\
\tvs
\frac{-187}{45} & \frac{2N}{15}-\frac{31}{81} & \frac{-377}{81} &
\frac{2N}{15}-\frac{167f}{405} & 0 & \frac{-167f}{405} \\
\tvs
\frac{2N}{15} & \frac{-377}{81} & \frac{2fN}{15}-\frac{754}{81} &
\frac{4N}{15}-\frac{2017f}{405} & 0 & \frac{-334f}{405} \\
\tvs
\frac{-35}{9} & \frac{2N}{15}-\frac{167f}{405} & \frac{4N}{15}-
\frac{2017f}{405} & \frac{2fN}{15}-\frac{179f^2}{405}-\frac{70}{9} &
0 & \frac{-179f^2}{405} \\
\tvs
0 & 0 & 0 & 0 & \frac{2fN}{15} & \frac{35f}{9} \\
\tvs
0 & \frac{-167f}{405} & \frac{-334f}{405} & \frac{-179f^2}{405} &
\frac{35f}{9} & \frac{187fN}{45}-\frac{179f^2}{405}
\end{array} \right)
\end{displaymath}

\begin{displaymath}
\hspace{-1.5cm}
D^{HV} = N C_f \!
\left( \begin{array}{cccccc}
\frac{2N}{15} & \frac{-1321}{315} & \frac{2N}{15} & \frac{-1321}{315} &0&0 \\
\tvs
\frac{-1321}{315} & \frac{2N}{15}-\frac{263}{567} & \frac{-14519}{2835} &
\frac{2N}{15}-\frac{263f}{567} & \frac{4}{945} &
\frac{-N}{105}-\frac{263f}{567} \\
\tvs
\frac{2N}{15} & \frac{-14519}{2835} & \frac{2fN}{15}-\frac{29038}{2835} &
\frac{4N}{15}-\frac{14519f}{2835} & \frac{8}{945} &
\frac{-2N}{105}-\frac{526f}{567} \\
\tvs
\frac{-1321}{315} & \frac{2N}{15}-\frac{263f}{567} & \frac{4N}{15}-
\frac{14519f}{2835} & \frac{2fN}{15}-\frac{263f^2}{567}-\frac{2642}{315} &
\frac{4f}{945} & \frac{-fN}{105}-\frac{263f^2}{567}\\
\tvs
0 & \frac{4}{945} & \frac{8}{945} & \frac{4f}{945} &
\frac{2fN}{15}-\frac{4}{105} & \frac{-2N}{105}+\frac{4387f}{945} \\
\tvs
0 & \frac{-N}{105}-\frac{263f}{567} & \frac{-2N}{105}-\frac{526f}{567} &
\frac{-fN}{105}-\frac{263f^2}{567} & \frac{-2N}{105}+\frac{4387f}{945} &
\frac{1433fN}{315}-\frac{263f^2}{567}-\frac{4}{105}
\end{array} \right)
\label{eq:A.1}
\end{displaymath}

\newpage



\clearpage
\noindent
\begin{center}
{\LARGE\bf Figures}
\end{center}

\begin{figure}[h]
\vspace{0.2in}
\centerline{
\epsfysize=1.5in
\epsffile{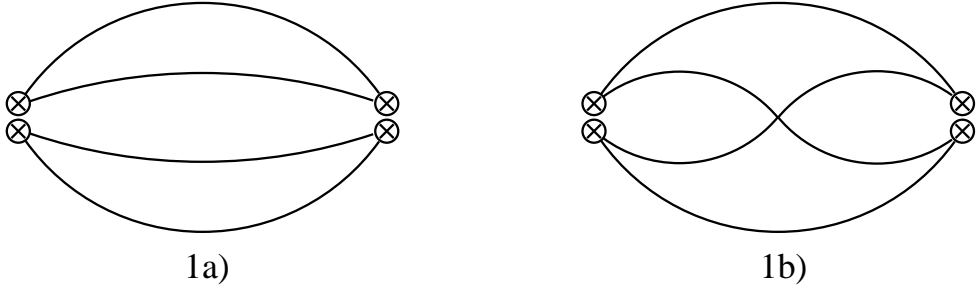}
\vspace{.2in}
}
\caption[]{Leading order diagrams for the 2--point function.\label{fig:1}}
\end{figure}

\setcounter{figure}{2}
\begin{figure}[h]
\vspace{0.2in}
\centerline{
\rotate[r]{
\epsfysize=4.5in
\epsffile{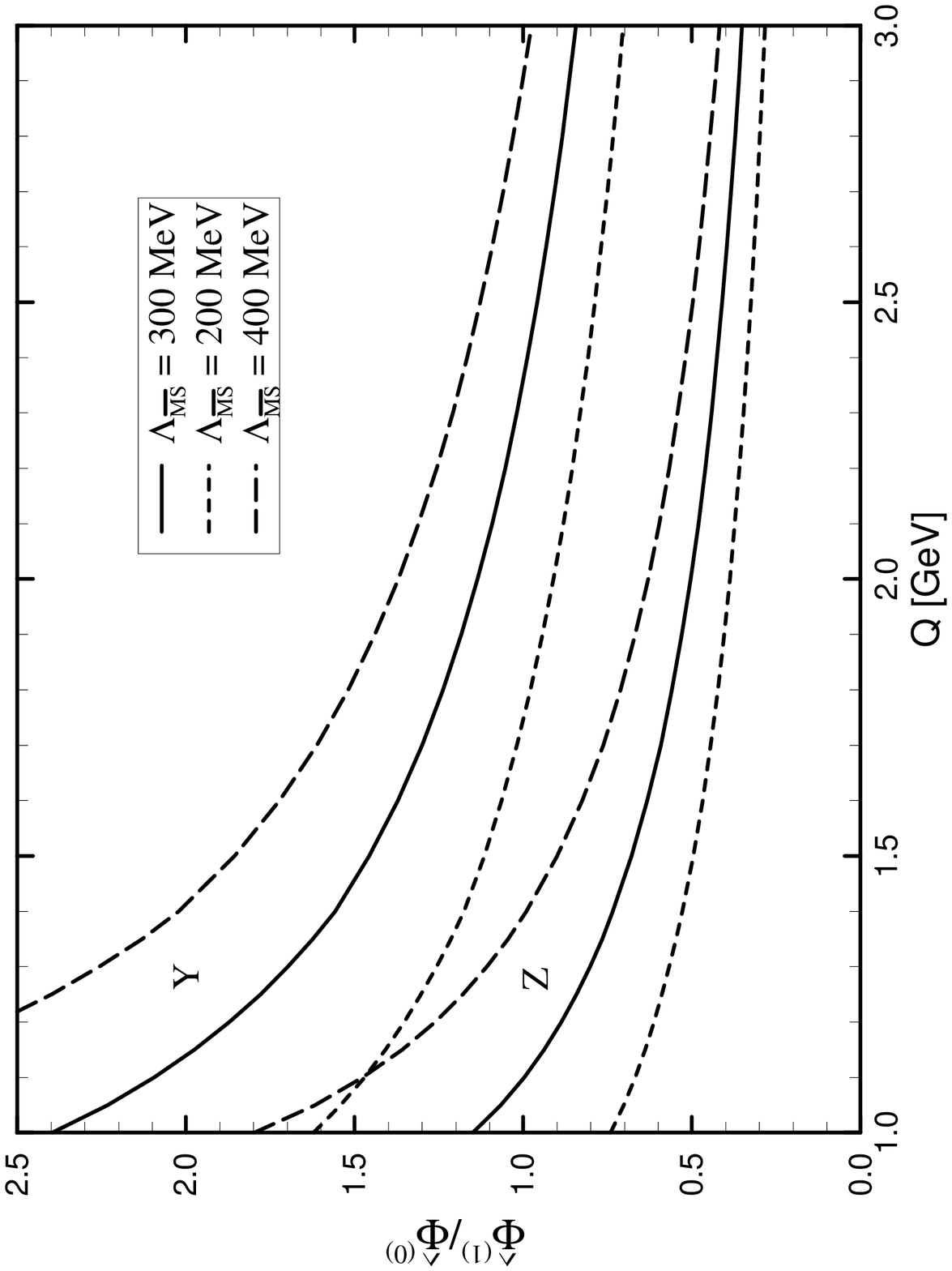}
\vspace{.2in}
}}
\caption[]{The ratios $\wh\Phi^{(1)}_{z}/\wh\Phi^{(0)}_{z}$ and
$\wh\Phi^{(1)}_{y}/\wh\Phi^{(0)}_{y}$.\label{fig:3}}
\end{figure}

\clearpage

\setcounter{figure}{1}
\begin{figure}[h]
\centerline{
\epsfysize=8in
\epsffile{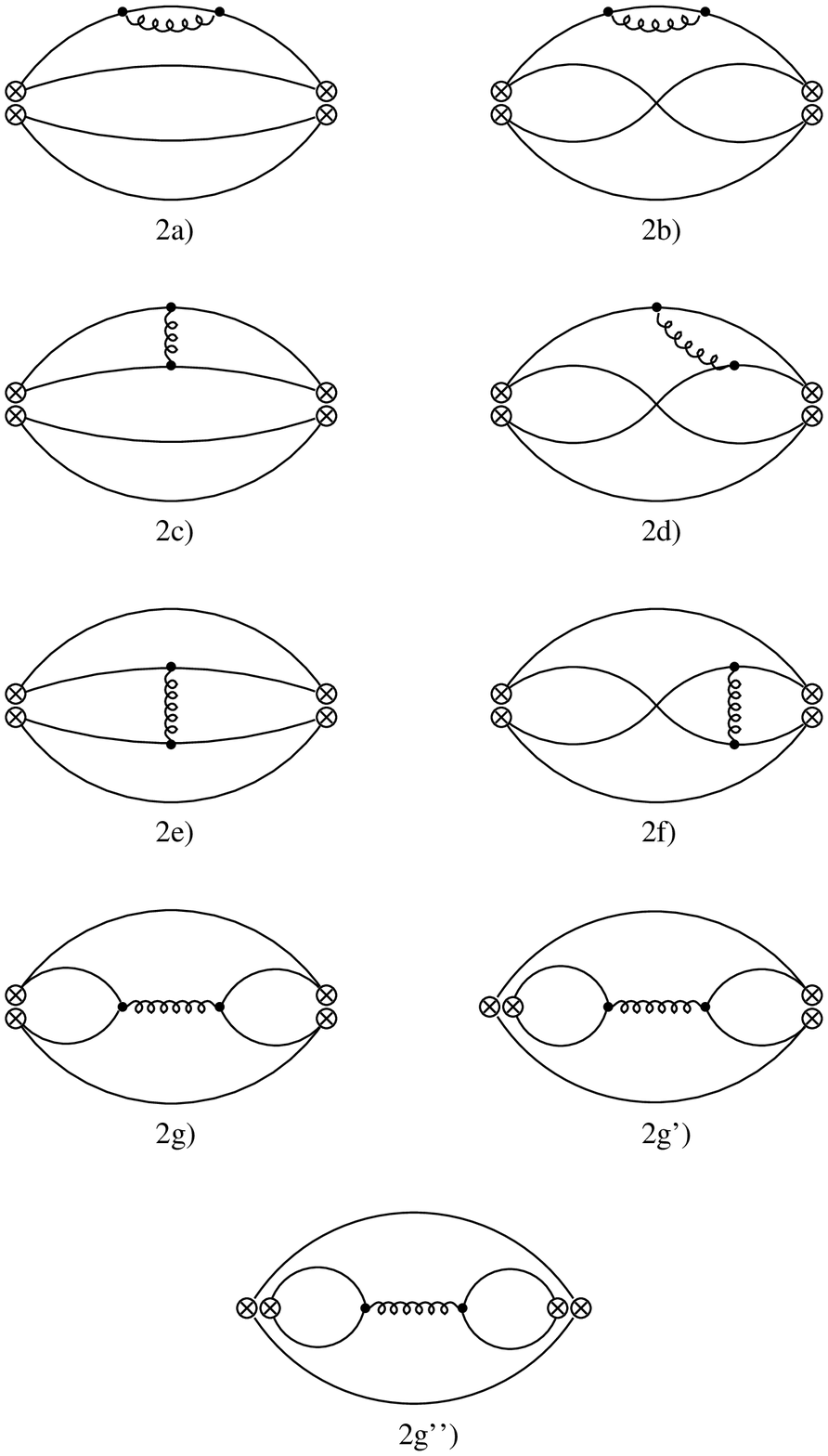}
\vspace{.1in}
}
\caption[]{$\cO(\al)$ diagrams for the 2--point function.\label{fig:2}}
\end{figure}

\end{document}